# Scheduling of Dependent Tasks Application using Random Search Technique

Deepak C. Vegda[*a] , Harshad B. Prajapati [b]
[a] Department of Information Technology, Dharmsinh Desai University, Nadiad-387001-Gujarat, India.
[b] Department of Information Technology, Faculty of Technology, Dharmsinh Desai University, Nadiad-387001-Gujarat, India
[*a] Email:- deepak.c.vegda@gmail.com.

*Abstract –*

*Since beginning of Grid computing, scheduling of dependent tasks application has attracted attention of researchers due to NP-Complete nature of the problem. In Grid environment, scheduling is deciding about assignment of tasks to available resources. Scheduling in Grid is challenging when the tasks have dependencies and resources are heterogeneous. The main objective in scheduling of dependent tasks is minimizing make-span. Due to NP-complete nature of scheduling problem, exact solutions cannot generate schedule efficiently. Therefore, researchers apply heuristic or random search techniques to get optimal or near to optimal solution of such problems. In this paper, we show how Genetic Algorithm can be used to solve dependent task scheduling problem. We describe how initial population can be generated using random assignment and height based approaches. We also present design of crossover and mutation operators to enable scheduling of dependent tasks application without violating dependency constraints. For implementation of GA based scheduling, we explore and analyze SimGrid and GridSim simulation toolkits. From results, we found that SimGrid is suitable, as it has support of SimDag API for DAG applications. We found that GA based approach can generate schedule for dependent tasks application in reasonable time while trying to minimize make-span.*

**1. Introduction**
Scheduling is a problem of assigning tasks to machine to complete their work. In this problem, we have a number of tasks and a number of resources. In general terms, scheduling is deciding about assigning jobs to workers. To allocate task to any machine we apply some heuristic or some logic. Because of its NP-complete nature scheduling is always a challenging problem whether it is in Grid or in any other environment. It requires more attention when tasks are dependent. As it is an NP-complete problem researchers are trying to apply heuristics to solve it. The goal is to get a solution which is near to the optimum solution. There is no standard heuristic for the group of NP-complete problems. It may be the case that one heuristic works well in some specific situation and not for other situation.

In this paper, we provide genetic algorithm based solution to solve dependent tasks scheduling problem. In our proposed work, we use height based approach with randomization technique to generate initial solutions. We define crossover and mutation operators which preserve dependency between tasks. We also consider data transfer among the dependent tasks. To respect dependency while performing mutation, we use height of each task. Our proposed algorithm is generalized and it is applicable to all dependent tasks scientific applications.

To test our proposed genetic algorithm we use simulator. We provide analysis of various available simulators for grid-environment in Table 3. We perform our experiments on SimGrid toolkit as its SimDag API suits our scheduling problem perfectly. In this toolkit DAG is prepared as an XML file, which is known as deployment file which defines our tasks graph. It represents our tasks and their dependency. There is one another file named platform file which represents our environment. For every run, we log the solution in a text file. We compare our proposed algorithm with min-min heuristic algorithm with respect to make-span. Our input to executable file of our algorithm is deployment file and platform file. The output is the solution of scheduling problem in form of information which shows tasks and their executions on particular machines with data transfer time. The main part of output will be the total make-span

To implement our proposed genetic algorithm in SimGrid toolkit, we use C language with SimDag API of SimGrid. We create following functions: *generatepopulation* to generate initial solutions, *get_height* to assign height to every task, *adjust_height* which adjusts the height of tasks to allow all the order of tasks, *crossover* and *mutation* for reproduction, *update_population* to update the available solutions. We take different DAGs to test our algorithm. One type of DAG has large execution time and low data transfer, second has large data transfer rate, in third type of DAG we have large number of task which may be run parallel. Our comparison criterion is the total make-span of an application.

This paper is organized as follow. We discuss about grid environment and the dependent tasks scheduling problem in section 3. Next, we discuss basic steps of genetic algorithm in section 3. Next, we put our proposed algorithm and its detail work in section 4 and finally shown the simulation and results in section 5. Finally, we conclude our work in section 6.

## 2. Related work

As scheduling is an NP-complete problem researchers try to apply heuristics or meta-heuristics to get optimum or near to optimum solution. People use a single heuristic or a combination of heuristics and meta-heuristics. It is called the hybrid meta-heuristics.

Edwin S.H, Nirwan Ansari and Hong Ren[1] have applied Genetic algorithm to multiprocessor scheduling. They have discussed the representation techniques for GA. They have also discussed a height based approach to validate the crossover and mutation operation for dependent tasks scheduling problem. They have compared their results with List algorithm. From this paper we use equation to find height of task.

A. Tamilarasi and T. Anantha kumar[2] have proposed combination of Simulated Annealing and Genetic algorithm for job-shop scheduling problem. Hadis Heidari and Abdolha chalechale[3] have also given the good representation techniques for genetic algorithm. Performance measure of this algorithm is to minimize the total completion time (make-span). They have compared their proposed work with Goncalves *et al* (2005) with taking cpu-time as performance measure.

Siriluck[5] has applied an Ant Colony Optimization for Dynamic Job Scheduling in Grid Environment. They compare their work with the existing simple algorithms like EDF and FCFS and the algorithm gives good results. They have compared total tardiness time and total schedule time.

Geoffrey Falzon and Maozhen Li [7] have used GA for dependent tasks scheduling problem. To preserve dependency in crossover, they discussed a crossover technique which changes the order of tasks. In crossover, they change the order of genes in offspring according to the second chromosomes. They have provided comparison of different selection methods and different crossover points. They conclude that ranked-based selection method obtains good output.

P Visalakshi and A.Bhuvaneswari[4] have applied Particle Swarm Optimization(PSO) and Simulated annealing for scheduling in multiprocessor environment. They have made a hybrid algorithm using PSO and SA. The concept of their proposed algorithm was to reduce the value of inertia weight dynamically. They have compared their algorithm with a simple PSO algorithm which has fix inertia weight.

Vincenzo Di Martino and Marco Mililotti[15] have used genetic algorithm to solve scheduling problem. Their objective was to maximize the throughput. Vincenzo Di Martino [16] has applied GA to find sub optimum solution. Jose Fernando Gonçalves and Jorge [17] have applied hybrid GA for job shop scheduling problem. Their objective was to improve total time for scheduling.

## 3. Background Theory

### 3.1 Grid Scheduling

Grid computing is the heterogeneous environment. In grid, request for resource is done by client and resource broker finds best suitable resource from the grid. Figure 1 shows simple structure of Grid environment.

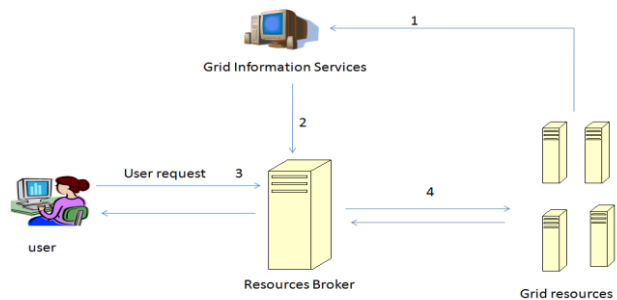

Figure 1. Grid Environment

1. In grid environment, information about the available resources and about their attributes is sent to GIS (Grid Information Service).

2. Resource broker retrieves this information from GSI.

3. Resource broker will find the best suitable resource or resources for the requests and allocate it to the user.

*Scheduling problem*: Suppose there are n numbers of dependent tasks (T1, T2, T3… Tn) and m number of resources (R1, R2, R3… Rm). Scheduling problem is stated that find a best suitable resource for each task to complete its execution. When tasks are dependent child task can't start its execution until all parent tasks have finished their execution.

Tasks are represented using DAG (Directed Acyclic Graph). It is defined as G = (V, E), where V is a set of v nodes/vertices and E is a set of e directed edges. Here nodes of DAG represent tasks and edges represent the dependencies between the tasks with directions. The source node of an edge is called the parent node while the destination node is called the child node. A node with no parent is called an entry node and a node with no child is called an exit [18]. DAG starts with a root node and ends with an end node. In figure 2, node1 is root node and node 10 is end node. Figure 2 also shows data which is transferred from the parent to their child. Using size of data and bandwidth between the machines we can calculate the communication cost. From figure 2, we can say that tasks 5 cannot start its execution until task 2 and task 6 completes their execution.

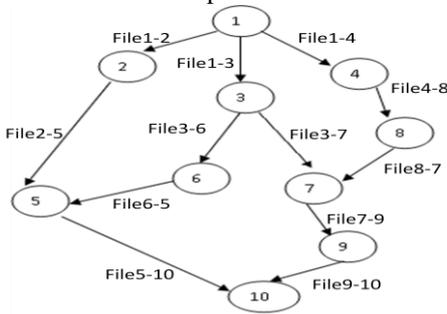

Figure 2 DAG with data transfer

Execution time of each task is listed below in table 1.

Table 1 execution time of tasks

| Task | 1 | 2 | 3 | 4 | 5 | 6 | 7 | 8 | 9 | 10 |
| --- | --- | --- | --- | --- | --- | --- | --- | --- | --- | --- |
| Exe-time | 21 | 12 | 18 | 12 | 9 | 21 | 15 | 24 | 11 | 10 |

### 3.2 Introduction to Genetic Algorithm

Genetic Algorithm (GA) is one of the meta-heuristics techniques. GA is based on the principles first laid down by Charles Darwin "survival of the fittest". Basic work of genetic algorithm is shown below in a figure 3.

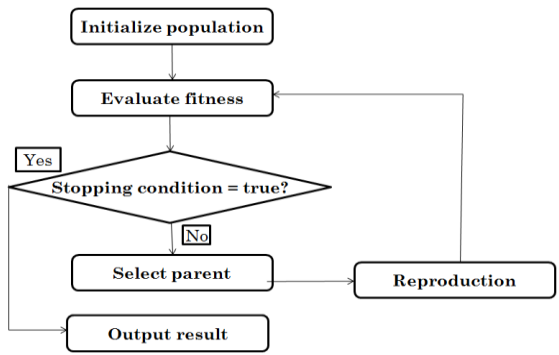

Figure 3 Basic Flow chart of Genetic algorithm

In first step, we randomly initialize some number of solutions which is known as population. Each solution is known as a chromosome. Each chromosome has its fitness value.

In second step, we calculate the fitness for each solution.

Next step is to check whether the stopping condition is achieved or not. In selection step, we select number of couples for producing new generation. We can use any available selection method or can randomly choose the couple.

Next step is to apply GA operators to selected parents and create new children. Crossover and mutation are the GA operations.

In crossover, we swap a selected portion or bunch of genes from a couple of chromosomes to each other and generate new solutions. The swapped portion is known as offspring. In mutation, we randomly select two points in a chromosome and swap their place. After applying GA operations now we have some new solutions which are actually generated using available solutions so we named them as children.

Now for new generated solutions, calculate the fitness value. Compare the fitness of available population and new generated children and update the population using the law of survival of the fittest. Remove the bad solutions from the population and replace better new children in population.

Repeat these steps until the stopping condition is true. Stopping condition may be the number of iteration or there is no change in the fitness value for some number of iteration.

## 4. Proposed Algorithm

We propose a genetic algorithm for scheduling of dependent tasks. It uses randomize height based approach. Below we have listed the steps of our proposed GA for dependent tasks scheduling.

> *Find height for each task in DAG*
> *Generate population*
> *Repeat*
> *Find the fitness for each solution*
> *Apply GA operators*
> *Update population*
> *Until stopping condition*

*4.1 Find height for each task in DAG*

To get the height of each task, we use a simple logic in DAG that if the task is child of a task then its height is 1+ height of its parent. Below equation is used to find the height [1].

> *Height [task] = 1 (if task is root)*
> **Otherwise**
> *Height [task] = max (height (parents)) +1*

Calculation of height starts from entry node and ends at exit node. After applying this equation, the height for each task in our DAG example is shown below in figure 4.

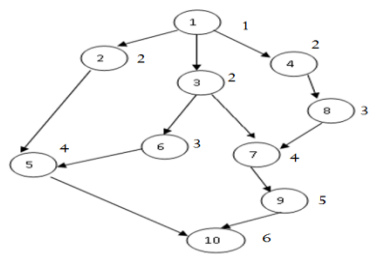

Figure 4 Height of tasks in DAG

Table 2 shows the height of each task in DAG.

Table 2 height of tasks

| Task   | 1 | 2 | 3 | 4 | 5 | 6 | 7 | 8 | 9 | 10 |
|--------|---|---|---|---|---|---|---|---|---|----|
| Height | 1 | 2 | 2 | 2 | 4 | 3 | 4 | 3 | 5 | 6  |

*4.2 Generate population*

To generate population perform the steps shown below:

> *Repeat*
> *For i=0 to number_of_task*
> *If (height [task[i]] = 1)*

*readytask []= task[i]*
*end if*
*end for loop*
*Generate a random number p*
*Select readytask [p]*
*Generate a random number r*
*Select machine [r]*
*adjust_height()*
*Repeat until there is no task with height 1*

*4.3 Adjust Height*

Adjust height function allows all possible order of the tasks. In this step, we update the height of the selected task as 0 and then update the height of the DAG again with this change. When we adjust the height, it updates the heights of tasks which are dependent on the selected task. This change is limited to this population only. For next generation, we will use our global height.

Suppose task 1 is selected first as its height is 1, so now its height for example shown in figure 4, is updated as 0. The tasks dependent on it will also make change in their height with respect to task 1. The updated height is shown in table 3 below.

Table 3 adjusted height of tasks in DAG

| Task   | 1 | 2 | 3 | 4 | 5 | 6 | 7 | 8 | 9 | 10 |
|--------|---|---|---|---|---|---|---|---|---|----|
| Height | 0 | 1 | 1 | 1 | 3 | 2 | 3 | 2 | 4 | 5  |

The advantage of this technique is it will allow all the possible order with dependency validation. After selection of each task for solution we apply this operation.

We also use a standard algorithm which assign machine to task based on load on that machine to generate the initial population. As we try to take benefit from the parent solutions, standard solutions help our algorithm to get better solution.

Representation of our solution is shown below in figure 5.

| M1    | M2    | M4    | M7    | M3    | M1    | M5    | M6    | M3    | M4     |
|-------|-------|-------|-------|-------|-------|-------|-------|-------|--------|
| Task1 | Task3 | Task6 | Task2 | Task5 | Task4 | Task8 | Task7 | Task9 | Task10 |

Figure 5 solution representation

*4.4 Fitness calculation*

Fitness value is the measure based on which we decide the fitness of solution. We measure the fitness of solution in form of make-span of that solution. The solution with minimum make-span is the fittest solution.

We are considering the dependency between the tasks as well as communication-cost between machines. A task may have more than one parent. We consider the maximum completion time of a parent to start execution of a task.

In example shown in figure 5, task5 have two parents, task2 and task6. When we calculate starting time of task5, we must consider their completion times on their regarding machines. In Figure 5, if task6 completes its work on machine4 at 5 unit and task2 completes at unit 7 on machine7 then execution of task5 will start from 7 on machine3 if no task is active at that time.

We derive a formula to calculate starting time of a task which is show below.

$$starting\_time = max\ (max\ (completion\ time\ of\ parents),\ available\ time\ of\ machine)$$

The completion time of that task is summation of starting time and its execution time. Its completion time is the available time for a machine on which it is executed.

$$completion\_time(task) = starting\_time\ (task) + execution\_time(\ on\ particular\ machine)$$

*4.5 Reproduction*

For reproduction, we select pairs of available solutions. To select solutions for reproduction we use a selection method. The selection method selects number of solutions to generate new populations. We use ranked based selection in our proposed algorithm because ranked base selection is better for this type of problems [7]. We assign ranks to each solution based on their fitness value and then select some pairs of solution and apply crossover and mutation to get new solutions. When we apply GA operations we take care about dependency. We do this in two ways, as discusses next.

In first way, we swap only the machine number and keep the task order as it is. Suppose, we have a pair of solution and we apply GA operators to generate new solution from them. To do this, we randomly generate an integer number from which we will swap the offspring. Suppose it is 5 in our example, which is shown by the arrow in figure 6. It is called crossover point. We swap machine number from crossover point in the selected pair of chromosomes. The output of this operation is shown in figure 7 and figure 8.

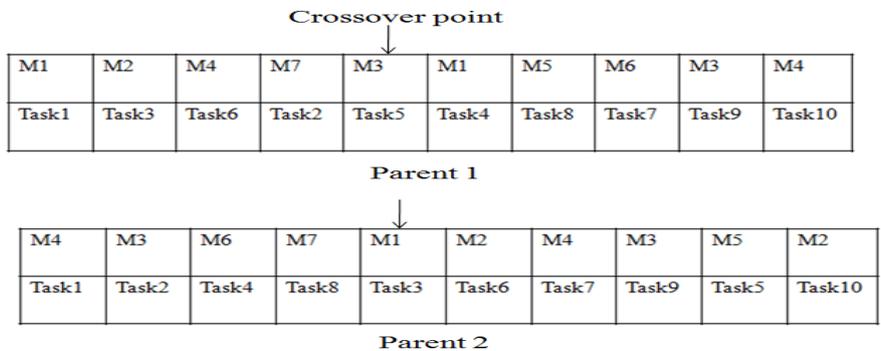

Figure 6 pair of solution for reproduction

Figure 7 child1 after crossover

Figure 8 child2 after crossover

The new generations preserve the dependency as we are not replacing the tasks of the chromosomes.

In second way, we randomly generate a crossover point and then swap the machine number of the tasks after crossover point. Take the above parents and crossover point for example; machine for task5 in parent1 and parent2 is swapped. This process continues until we complete swapping for all the tasks which are after crossover point in both solutions. The new solutions after this crossover process is shown below in figure 9 and figure 10.

Figure 9 child3 after crossover

Figure 10 child4 after crossover

For mutation, we just simply generate two points randomly and make a check whether there is a dependency between tasks at those points or not. If no dependency presents then we swap their position with machine number and if yes then we generate another point which allow mutation. Suppose, mutation points for parent1 in figure 6 are 3 and 4. Now we check whether task6 and task2 are dependent or not. There is no dependency between them so we swap them and generate new solution. New solution after mutation is shown below in figure 11.

| M1 | M1 | M7 | M4 | M3 | M1 | M5 | M6 | M3 | M4 |
|---|---|---|---|---|---|---|---|---|---|
| Task1 | Task3 | Task2 | Task6 | Task5 | Task4 | Task8 | Task7 | Task9 | Task10 |

Figure 11 new solution after mutation

After reproduction method, we calculate fitness value for new child and compare it with available solution. We remove bad solutions which are in current population and place good solutions from new generation in place of them. We repeat these steps until we reach the stopping condition. The stopping condition is number of iteration or no change in fitness value for some number of iterations.

## 5. Implementation of Proposed Algorithm

### 5.1 Evaluation of available simulators for Grid Scheduling

There are two ways to test our proposed algorithm. One is do it on real grid environment and second is to use simulator. It is very difficult to create a real environment and configure it. This process takes time. It is easy to use simulator to test our proposed work.

The main aim of simulator is to test the implementation work in the absence of the required environment. To perform the simulation, we use SimGrid toolkit which allows us to compare the scheduling algorithm for grid environment. We provide some analysis on available toolkits. We provide a comparison between toolkits.

Table 4 comparison of simulation toolkit for Grid Scheduling [10]

| Toolkit | Programming-Language | Scale | Domain | Orientation | Extension | Scheduler | Task dependency |
|---|---|---|---|---|---|---|---|
| SimGrid | C, JAVA | Few 10,000 | Grid, D.E, P2P, Cloud | Event | Data grid, Network | Static, dynamic | Allow |
| GridSim | JAVA | Few 100 | Grid, Cloud | Event, Process | Data grid, Network | Static, dynamic | Allow |
| OptorSim | JAVA | Few 100 | (Data) Grid | Data Grid | Network | Static | No |

### 5.2 Architecture of SimGrid

SimGrid has six main components. These components are SimDag, MSG, GRAS, SMPI, XBT and SURF. In this SimDag, MSG, GRAS, SMPI are API of SimGrid toolkits.

SimDag is used for Comparing Scheduling algorithm for DAG applications. Our purpose is also to check the scheduling algorithm for DAG applications. So, we will use SimDag API for our experiment purpose. For our research SimDag is the perfect API as it has all the functionality related to DAG. We present our task graph in form of DAG, and SimDag gives Programming environment for DAG applications.

We use SimDag programming environment of the SimGrid because we want to compare algorithms and heuristics with DAGs of parallel tasks [13]. In SimDag we have data types for tasks, workstations, task dependency and etc. There are functions available which help us to implement our proposed algorithm. Some functions and their work is shown below:

- *SD_daxloadar():* load the deployment file and return the all the tasks. We get tasks from the XML file using this function.
- *SD_task_get_parents(task)* : return the parents of the desired task.
- Same way we can find children of the task using function *SD_task_get_children(task).*
- *SD_task_get_name(task)* returns the name of the task.

### 5.3 Implementation of our proposed algorithm

For implementation of our proposed genetic algorithm, we choose C programming language with SimDag API of SimGrid. To store initial solutions, we create a structure named chromosome. Its definition is shown below in figure 12.

```
struct chromosome
{
    SD_workstation_t cmachine[100];
    SD_task_t ctask[100];
    double cfitness;
}
```

Figure 12 declaration of structure for chromosomes

Here *cmachine* array is to store machines for tasks, *ctasks* is to store the order of tasks for each generation and *cfitness* is to store the fitness of the solution.

To store new generation, which is created using reproduction method, we define another structure named *newgeneration*, which is shown below.

```
struct newgeneration
{
    SD_workstation_t nmachine[100];
    SD_task_t ntask[100];
    double nfitness;
}
```

Figure 13 declaration of structure for new generation

To load DAG (task graph) in a dynamic array, we use *SD_daxload* function, which is inbuilt in SimDag. After this, we find height for each task using the equation we have defined in section 4. We define *Get_height* function to do this, which returns array of height containing height for each task.

We perform *Generate_population* function which calls the functions which generate order of tasks for each generation and randomly selects machine from the environment for each tasks. We define *Calculate_fitness* function to calculate fitness value (make-span) for each generation.

For each solution order of tasks is stored in *ctask* array, order of machines is stored in *cmachine* array and fitness value will be stored in *cfitness*.

We provide *selection_method* function to select number of pair for new generation based on their fitness value. On the pair of solutions, we perform crossover and mutation operation as we discussed in section 4. We store new generation's order of tasks and order of machines in *ntask* and *nmachine* respectively which are the variables of *newgenerstion* structure.

We also provide a function *update_populationl*, which replaces bad solutions in population comparing their fitness values. The rule Survival of the fittest will preserve here. We continue these steps until stopping condition is reached.

*5.4 Results and Comparison*

It is easy to compile and run our algorithm in SimGrid on windows platform. To compile code, just perform the steps listed below:
1. Run windows shell "cmd".
2. Open your project Directory ('cd' command line).
3. Create a build directory and change directory. (optional)
4. Type 'cmake -G"MinGW Makefiles" <path_to_your_project>'
5. Run mingw32-make

Figure 14 compile the project in SimGrid

To run the project, write the name of executable file with platform and deployment files as input.

Figure 15 run a project in SimGrid

We log the output of our algorithm in a text file which is shown below in figure 16.

```
2   Schedule ID00001@job1 on Machine2
3   Schedule ID00002@job2 on Machine2
4   Schedule ID00004@job4 on Machine1
5   Schedule ID00003@job3 on Machine2
6   Schedule ID00008@job8 on Machine2
7   Schedule ID00006@job6 on Machine1
8   Schedule ID00007@job7 on Machine2
9   Schedule ID00005@job5 on Machine1
10  Schedule ID00009@job9 on Machine2
11  Schedule ID00010@job10 on Machine2
12  Simulation Time: 74.120253
```

Figure 16 output in a text file

Table 5 comparison of proposed algorithm with min-min

| No. of tasks/machines | No. of parallel tasks | Proposed GA (make-span) | min-min (make-span) |
| --- | --- | --- | --- |
| 10 / 2 | 3 | 70.12 | 74.12 |
| 10 / 7 | 3 | 102.94 | 105.41 |
| 25 / 2 | 10 | 95.38 | 98.21 |
| 25 / 7 | 10 | 94.55 | 98.21 |
| 45 / 2 | 7 | 196.45 | 201.96 |
| 45 / 7 | 7 | 235.45 | 239.72 |
| 2 / 90 | 0 | 489.31 | 515.19 |
| 10 / 90 | 3 | 2279.73 | 2280.04 |
| 40 / 90 | 10 | 3421.25 | 3850.68 |

In table 5, we show comparison between our proposed algorithm and existing min-min algorithm for dependent task scheduling problem. We compare performance of algorithm using make-span. Using maximum 100 initial solutions and maximum 50 iteration our algorithm produces less make-span than min-min. We take different task graphs and environments. It is fact that our proposed algorithm take more time than min-min.

**Conclusion**
We proposed a randomize height base GA to solve scheduling problem of dependent tasks. In this paper, we generate initial solutions which preserve the dependency. We also provide crossover and mutation with valid task order. We take 100 initial solutions and take 50 maximum iterations. We test our proposed algorithm on different tasks graph providing different environments. For each task graph our algorithm performs better than min-min and minimizes the make-span. Our proposed algorithm is comparably better than min-min in most of the situation. SimGrid toolkit is a good simulator to perform scheduling algorithm. In future we wish to add a new step in our proposed algorithm which will store offspring during the crossover and produce new solution using that offspring.

# References

[1] Edwin S.H.Hou, Nirwan Ansari, Hong Ren, A "Genetic-Algorithm for Multiprocessor Scheduling, IEEE Transactions on Parallel and Distributed systems, Vol.5, No.2.

[2] A.Tamilarasi and T.Anantha Kumar,(2010),"An enhanced genetic algorithm with simulated annealing for job-shop Scheduling",International Journal of Engineering, Science and Technology Vol 2, No 1,pp 144-151.

[3] Hadis Heidari and Abdolha chalechale, "Scheduling on Multiprocessors system using Genetic Algorithm", International Journal of Advanced Science and Technology Vol. 43, June, 2012

[4] S.N.Sivanandam, P Visalakshi, and A.Bhuvaneswari (2009) "Multiprocessor Scheduling Using Hybrid Particle Swarm Optimization with Dynamically Varying Inertia", international journal of computer science and applications, Vol. 4 Issue 3, pp 95-106

[5] Siriluck Lorpunmanee, Mohd Noor Sap, Abdul Hanan Abdullah, and Chai Chompoo-inwai, "An Ant Colony Optimization for Dynamic Job Scheduling in Grid Environment", International Journal of Computer and Information Engineering 1:8 2007

[6] Fr_ed_eric Sute (2010), "Simulating DAG scheduling algorithms with SimDAG".

[7] Geoffrey Falzon and Maozhen Li, "Enhancing genetic algorithms for dependent job scheduling in grid computing environments", J Supercomput (2012) 62:290–314 DOI 10.1007/s 11227-011-0721-2

[8] Martin Quinson, Arnaud Legrand, Henri Casanova, "The SimGrid Framework for Research on Large-Scale Distributed Systems".

[9] Rajkumar Buyya, and Manzur Murshed , "GridSim: a toolkit for the modeling and simulation of distributed resource management and scheduling for Grid computing"

[10] Martin Quinson, Arnaud Legrand, Henri Casanova "The SimGrid Framework for Research on Large-Scale Distributed Systems"

[11] Ian Foster, Carl kesselman(1999), "The GRID 2, blueprint for new computing infrastructure", Morgan Kauffmann.

[12] Min Liu*, Cheng Wu (2003) ,"Scheduling algorithm based on evolutionary computing in identical parallel machine production line" Robotics and Computer Integrated Manufacturing 19 401–407

[13] Da SimGrid Team(June 13, 2012), "Using SimGrid 101 Getting Started to Use SimGrid".

[14] Henri Casanova, "Simgrid: a Toolkit for the Simulation of Application Scheduling"

[15] Marco Mililotti, Vincenzo Di Martino(2002), "Scheduling in a Grid computing environment using Genetic Algorithms"0-7695-1573-8/02/ (C) IEEE

[16] Vincenzo Di Martino(2003), "Sub Optimal Scheduling in a Grid using Genetic Algorithms", 0-7695-1926-1/03/ (C) IEEE

[17] Jose Fernando Gonçalves , Jorge Jose de Magalhães Mendes , Maurício G. C. Resende(September 2002), "A Hybrid Genetic Algorithm for the Job Shop Scheduling Problem", AT&T Labs Research Technical Report TD-5EAL6J,.

[18] Dr. D.I. George Amalarethinam,G.J. Joyce Mary(April 2011), " A new DAG based Dynamic Task Scheduling Algorithm (DYTAS) for Multiprocessor Systems", International Journal of Computer Applications (0975 – 8887) Volume 19– No.8.